\documentclass[ajp,preprint]{revtex4-1}

\usepackage{amsmath}  
\usepackage{amsfonts} 
\usepackage{graphicx} 
\usepackage{subfigure} 

\begin{document}


\title{An Analytical Approach to \\Eddy Current in Electromagnetic Damping}

\author{Hao Chen}\thanks{First author}
\email{hao.chen.hc728@yale.edu; chenhao2@shanghaitech.edu.cn} 
\affiliation{Department of Physics, Yale University, P.O. Box 202736, New Haven, Connecticut 06520, USA}
\altaffiliation{School of Physical Science and Technology, ShanghaiTech University, Pudong, Shanghai 201210, China}

\author{Yawen Xiao}\thanks{First author}
\email{xiaoyw@shanghaitech.edu.cn}
\affiliation{Department of Physics, University of Illinois Urbana-Champaign, Urbana, Illinois 61801-3003, USA}
\altaffiliation{School of Physical Science and Technology, ShanghaiTech University, Pudong, Shanghai 201210, China}

\author{Jinze Shi}
\email{shijz@shanghaitech.edu.cn}
\affiliation{School of Physical Science and Technology, ShanghaiTech University, Pudong, Shanghai 201210, China}

\author{Weishi Wan}\thanks{Corresponding author}
\email{wanws@shanghaitech.edu.cn}
\affiliation{School of Physical Science and Technology, ShanghaiTech University, Pudong, Shanghai 201210, China}

\begin{abstract}
An analytical method of calculating eddy current in a metallic spinning gyroscope in external magnetic field is presented. With reasonable assumptions, the problem is simplified from the time-dependent one governed by Maxwell equations to the boundary value problem of Poisson's equation, which yields a closed-form expression of the eddy current. The rotation frequency as a function of time is calculated, compared with experiment and the relative error is found to be 8.61 \%\ .
\end{abstract}

\maketitle 

\section{Introduction} 

The International Young Physicist Tournament (2019) contains a problem of a spinning gyroscope made from a conducting, yet non-ferromagnetic material that slows down when placed in a magnetic field due to the interaction between induced eddy current and the magnetic field\cite{iypt}. Such a phenomenon, which is usually referred to as \'electromagnetic breaking' or \'electromagnetic damping', is widely discussed qualitatively in college and high school level electromagnetism textbooks as a direct result of Faraday's law of electromagnetic induction\cite{textbook}. Quantitatively, the eddy current in electromagnetic damping has also been studied rather extensively by physicists and electrical engineers over the past decades. W. R. Smythe calculated the eddy current in a conducting disk rotating in an external magnetic field generated by one cylidrical permenant magnetic pole on each side\cite{Smythe1942}. Later, another approach to this problem was developed by D. Schieber, which gave the result that was in agreement with Smythe's\cite{Schieber1974}. In both of the two papers, the Maxwell equations are solved in the frame of reference that moves together with the rotating disk. Recently,  M. A. Nurge et al. studied the distribution of eddy current in a rotating sphere in external magnetic field in the lab frame\cite{Nurge2018}. Although their derivation was rigorous, they expressed the result as a sum of a series rather than a closed-form formula, and each term of the series was a multiple of an associated Legendre polynomial. Such a result is rather complicated for application to a practical case. In addition, one has to expand the external magnetic field with associated Legendre polynomials in advance, making their result even more complicated. It is advantageous to develop a method of solving the eddy current problem in lab frame which gives a closed-form result that is easy to use.

In this paper, we present a general method of calculating eddy current based on solving the Maxwell equations with some reasonable assumptions, which can be further simplified to a boundary value problem of a Poisson's equation. In general cases, the solution of the Poisson's equation can only be expressed as a sum of a series. However, in order to have a better understanding of the underlying physical mechanism and make the result easy to be applied to practical cases, deriving a closed-form formula is of great significance, if achievable. Fortunately, in our gyroscope problem, the closed-form formula of eddy current can be obtained with a simplified expression of the magnetic field. Meanwhile, all the idealized conditions in our theory can be realized with a simple apparatus in experiments. While it is hard to directly measure eddy currents in conducting materials, which is the rotator of our gyroscope, it is straightforward to compare the deceleration processes deduced in theory and observed in experiment to verify our closed-form result of the eddy current.

In this article, to show the whole picture of our work, we firstly stated the basic logic of our theory of calculating eddy currents in a block of conducting material. Then, we apply the theory to the gyroscope problem to get a theoretical result for the deceleration process. The comparison of the theoretical and experimental results shows that they fit each other extremely well. Lastly, we discuss the justification of all the assumptions of our model and give a qualitative explanation of the remaining 8.61\%\ error.

\section{Basic logic of the theory and its application}
In this section, we will discuss how to apply electromagnetism laws to calculate eddy currents, which starts with Maxwell equations\cite{feynman}:
\begin{eqnarray}
\vec{\nabla}\cdot\vec{E} = \frac{\rho}{\varepsilon_{0}},\\
\vec\nabla\times\vec{E} = - \frac{\partial\vec{B}}{\partial t},\\
\vec\nabla\cdot\vec{B} = 0,\\
c^{2}\vec\nabla\times\vec{B} = \frac{\vec{j}}{\varepsilon_{0}} + \frac{\partial\vec{E}}{\partial t},
\end{eqnarray}
where ${\vec E, \vec B}$ are electric and magnetic fields, while ${\rho , \vec j }$ are the densities of charge and current, respectively. (We may call them 'charge' and 'current' in the rest of the article to be concise.)

Since we are dealing with current in a piece of conductive material, Ohm's law must be taken into account to describe it:
\begin{equation}
\vec{j} = \sigma(\vec{E} + \vec{f}),
\end{equation}
where ${\vec{f}}$ is the 'non-electrostatic force', which is usually inside batteries. Its loop integral along the circuit is 'electromotive force', usually denoted as 'emf'.\cite{griffiths}

In the problem of the gyroscope, the magnetic field is produced by permanent magnetic blocks and hence independent of time. (The magnetic field produced by eddy current is omitted as usual.) As a result, the electric field in the conducting material (rotator of the gyroscope) obeys the reduced form of Equation (1) \&\ (2):
\begin{eqnarray}
\vec\nabla\cdot\vec{E} = \frac{\rho}{\varepsilon_{0}},\\
\vec\nabla\times\vec{E} = 0.
\end{eqnarray}

To make the above equations easier to solve, it is convenient to set up a scalar field: ${\varphi (\vec{r})}$, which is called scalar potential in electromagnetism, whose minus gradient is the electric field (${-\vec\nabla{\varphi} = \vec{E}}$, and it satisfies (7) automatically). Plugging this into Equation (6), we have:
\begin{equation}
-\nabla^{2}\varphi = \frac{\rho}{\varepsilon_{0}}.
\end{equation}

Now, it appears that Equation (8), together with proper boundary conditions, is sufficient to determine the scalar potential ${\varphi}$ (and then electric field). However, one cannot calculate scalar potential ${\varphi}$ from Equation (8) yet, since we do not have a priori knowledge of the charge distribution. Therefore, a different perspective must be brought in to construct another relation between ${\varphi}$ and ${\rho}$, in order to get a set of solvable equations of scalar potential. It is worth noting that ${\vec f}$ together with ${\vec E}$ in Equation(5) drives the motion of charge, and with this, the conservation law of charge:
\begin{equation}
\vec\nabla\cdot\vec{j} = \frac{\partial\rho}{\partial t},
\end{equation}
can be used to establish the relationship between current and charge. By plugging Equation (5) into (9), and then replacing ${\vec E}$ with ${-\vec\nabla\varphi}$, another equation linking electric field and charge density is derived:
\begin{equation}
\sigma(-\nabla^{2} \varphi + \vec\nabla\cdot\vec{f}) = \frac{\partial\rho}{\partial t}.
\end{equation}

Then, by solving Equations (8) \&\ (10), a result of ${\varphi}$ and then ${\vec E}$ and finally current ${\vec{j}}$ can be obtained. Note that, in order to solve these equations, proper boundary and initial conditions are needed. Since initial conditions are unique and straightforward in different problems, we only list the boundary condition below, which is derived from the fact that the current cannot go through the surface of the material:	
\begin{equation}
\vec{j}\cdot\hat{n}\mid_{boundary} = j_{n}\mid_{boundary} = 0,
\end{equation}
where ${\hat{n}}$ is the unit normal vector of the boundary.

In summary, we construct a general method to calculate current in conducting material. For static magnetic field, the electric field has to be ``irrotational'' (${\vec\nabla\times\vec E = -\partial\vec B/\partial t = 0}$). Other parameters, such as the shape of the conductor, the conductivity or the distribution of non-electrostatic force can be arbitrary. While difficult, the second order time-depending partial differential equations below are solvable:
\begin{eqnarray}
\begin{cases}
-\nabla^{2}\varphi = \frac{\rho}{\varepsilon_{0}},\\
\sigma(-\nabla^{2} \varphi + \vec\nabla\cdot\vec{f}) = \frac{\partial\rho}{\partial t},\\
j_{n}\mid_{boundary} = 0.
\end{cases}
\end{eqnarray}
In order to solve these equations in our problem, we are going to introduce several assumptions to simplify the calculation (justification for these assumptions will be discussed in the last section of this article):
\begin{enumerate}
\item The current varies so slowly with time that it can be treated as nearly time-independent, which means ${\partial \rho / \partial t = 0}$.
\item The non-electrostatic force here is due to the rotation of the gyroscope, which can be expressed as:
\begin{equation}
\vec{f} = \vec{v}_{rot}\times\vec{B} = (\vec{\omega}\times\vec{r})\times\vec{B},
\end{equation}
where ${\vec{v}_{rot}}$ represents the velocity of a particular point on the rotator, ${\vec\omega}$ is the angular velocity of the rotator and ${\vec r}$ is the position vector from the center of the gyroscope.
\item The magnetic field supplies the current, and this force acts on the gyroscope directly causing its rotation slow down, but does not affect current distribution.
\end{enumerate}

With these assumptions, we can use a much more concise Poisson's Equation to calculate the current in the gyroscope:
\begin{equation}
\nabla^{2}\varphi = \vec\nabla\cdot[(\vec{\omega}\times\vec{r})\times\vec{B}].
\end{equation}
We then calculate the deceleration process by evaluating the magnetic torque on it.

We will show in the next section that the current has a rotational form and can be considered as 'eddy current'. Additionally, it is worth pointing out that such 'eddy current' is a common phenomenon when non-electrostatic force is originally produced by magnetic field (both magnetic force and the force given by induced electric field). Such 'eddy current' is harmful in many cases since it cause extra loss of energy.

\section{Modeling}

We turn to the simplest case to solve the problem analytically. Here we only take the component of the field perpendicular to the plane of the rotator into consideration. The in-plane magnetic field, whose role will be discussed in the last part of the article, has a much smaller contribution to eddy current due to the fact that the radius of the rotator is much larger than its thickness.

Qualitative analysis shows that there will be no stable current in the gyroscope placed in an even magnetic field because the non-electrostatic force distribution has a rotational symmetry (the free electrons will immediately form an electric field to cancel non-electrostatic force precisely). Therefore, a magnetic field distributed anti-symmetrically is the simplest form for us to setup the experiments and analyze the problem theoretically.

\begin{figure}[h!]
\centering
\subfigure[The schematic diagram of our model]{
\includegraphics[width=3in]{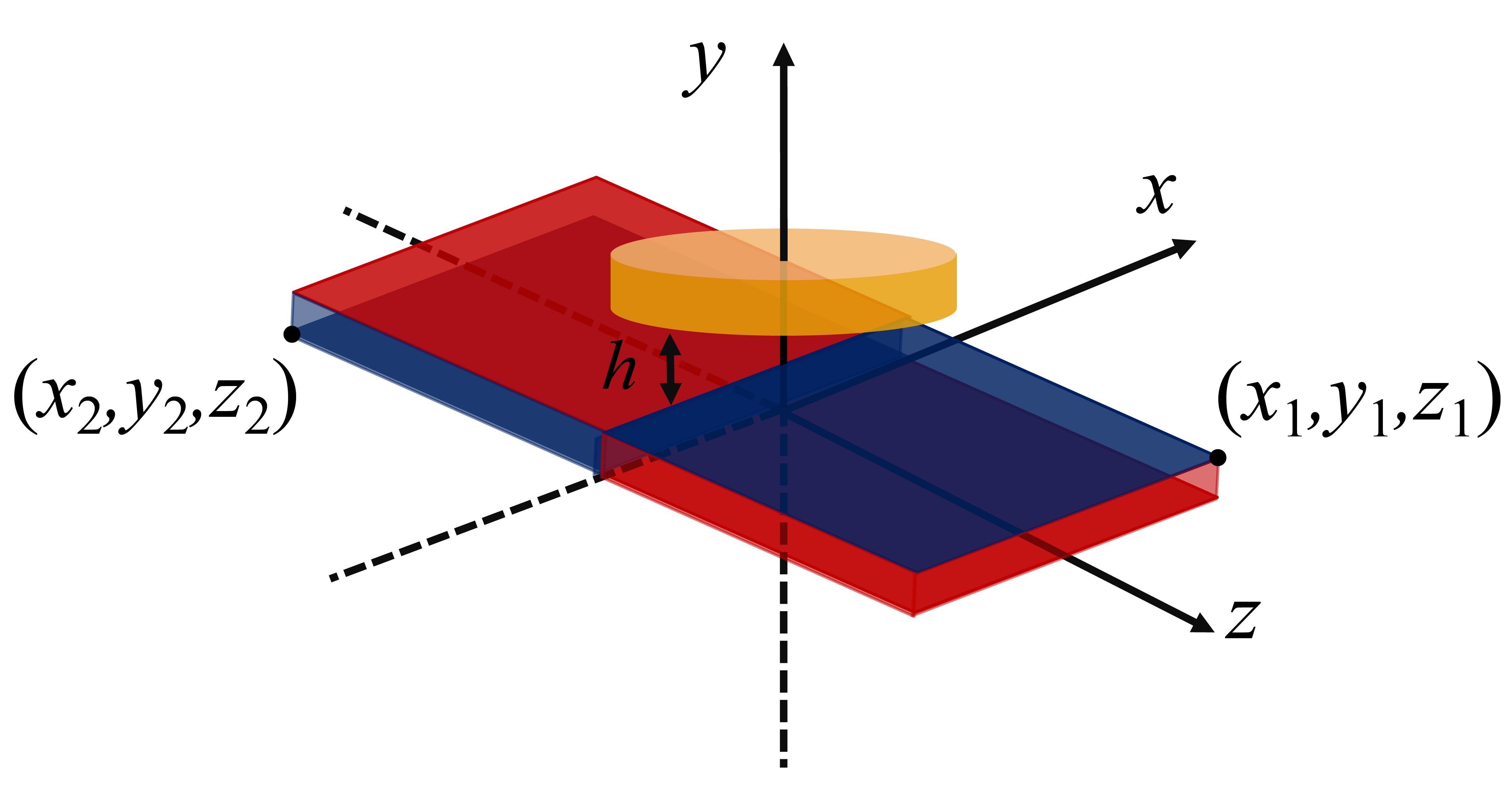}
}
\subfigure[Vertical component of the field]{
\includegraphics[width=3in]{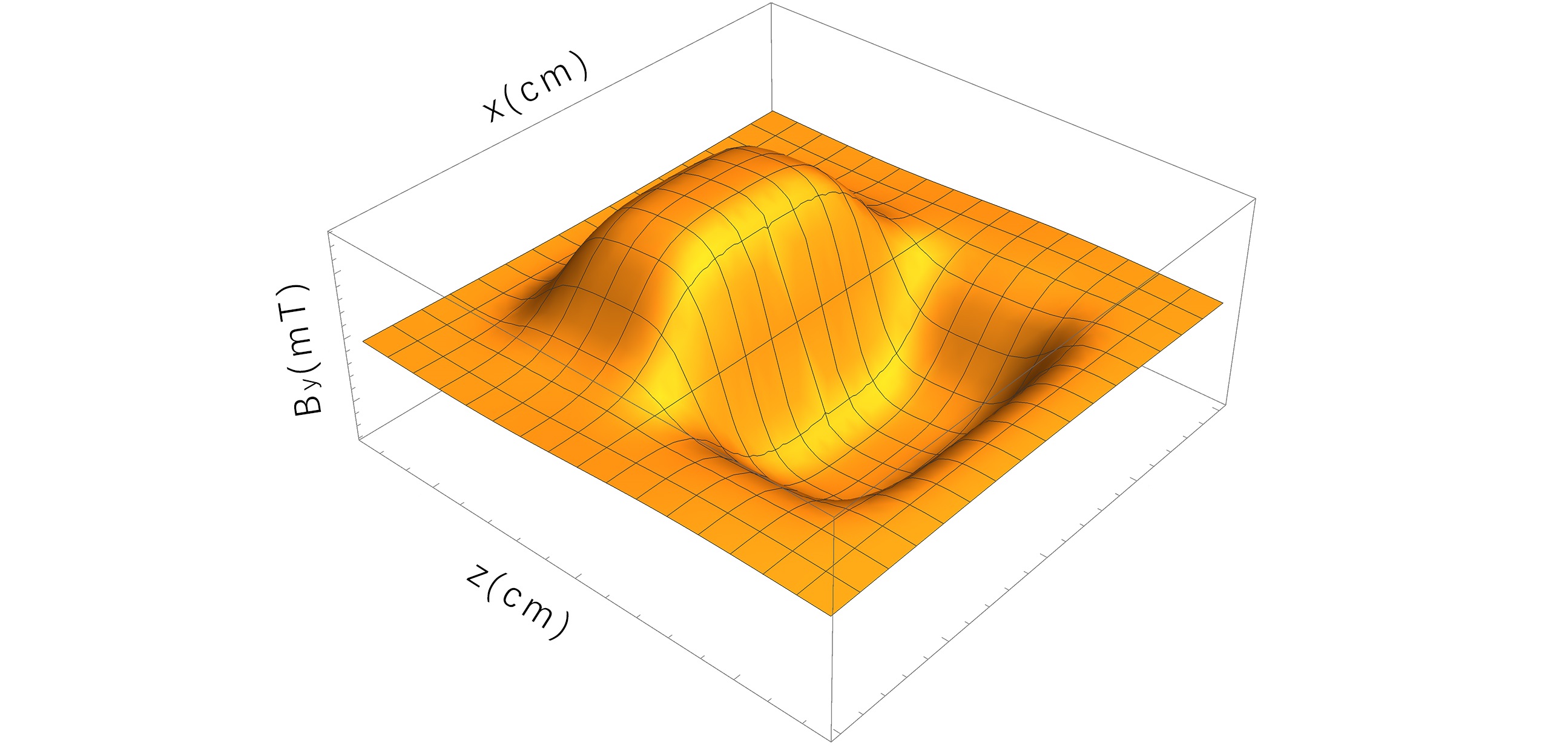}
}
\caption{Figure (a) gives a schematic view of our model. It shows the relative position between the magnets (red-blue blocks) and the rotator of our gyroscope (yellow cylinder). Red and blue parts represent the north and south poles of the magnets respectively. Figure (b) is a plot of the vertical component (y component) of the field at the height of the rotator. We find that the magnetic field is anti-symmetric about the x-y plane. (We will discuss the horizontal component in the last section, since it contributes to our rest 8.61\%\ error.)}
\end{figure}

To produce such an anti-symmetric field, we set two identical cuboid magnets beside each other with the north polarity of one matched to the south polarity of the other. When we do this the magnets attract and stick together. Such a configuration forms an anti-symmetrical magnetic field above it, in the vertical direction. (The picture of our model and the vertical component of field distribution is shown in FIG.1).The gyroscope is placed on the center line where the two magnets meet, with the gyroscope¡¯s brass rotator parallel to the magnetic surface. The center of the brass rotator is on the middle of the line between the two magnets. These equipments and experiment process are shown in FIG.2 and its illustration.

\begin{figure}[h!]
\centering
\includegraphics[width=3in]{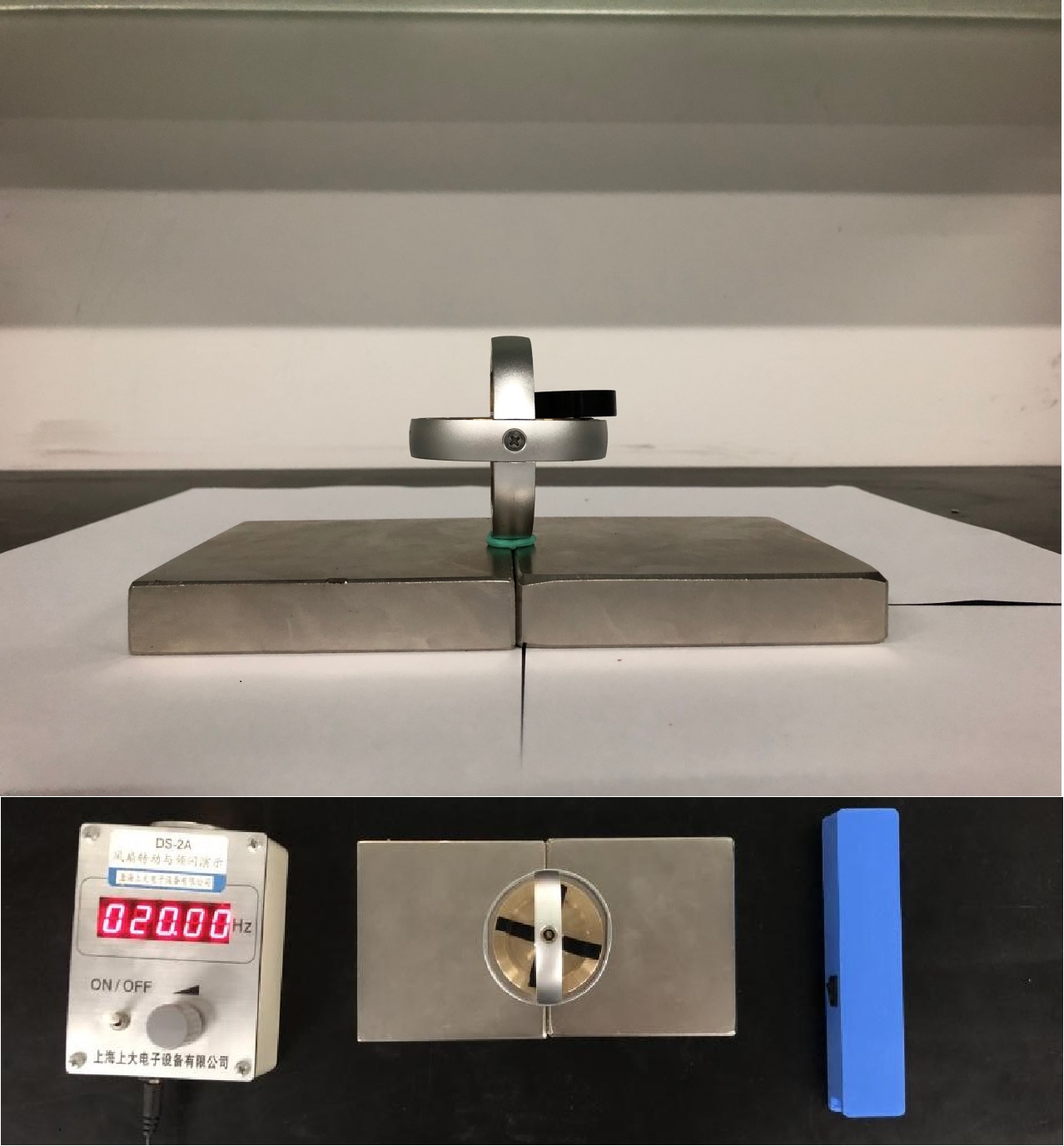}
\caption{Here shows how the experiment was conducted. The blue rectangular device consists of a small rotating motor, and we use the device to initialize the rotation of the gyroscope, as the magnetic field is not strong enough to start the rotation. Attached to the rotator are four identical pieces of tape, which form a cross that we use to measure the rotation of the gyroscope. Then, using a 20.00 Hz strobe light, we measure the angular velocity of the rotator during the process of deceleration. When the adhesive tape appears to rotate at a diminishing rate, then begins to rotate in the other direction, at the instance of the directional change, the frequency of the rotator has synchronized with the strobe light. When the strobe light synchronizes, the revolution of the rotator is an integral multiple of 5.000 r/sec (a quarter of 20.00 Hz). \\
Notice: 20.00 Hz was chosen as an appropriate frequency because higher frequencies make video recording difficult. Conversely, lower frequencies make us unable to catch the directional change with the strobe light.\\
Detailed information of our apparatus: 1. Magnets: 10.00cm * 10.00cm * 2.00cm; 2. Rotator: made of brass (type: H59), 5.30cm in diameter; 3. Distance between top surface of the magnets and the rotator: ${h=3.10cm}$.}
\label{Experimental devices}
\end{figure}

\section{Distribution of magnetic field}
Before dealing with the gyroscope problem, we first focus on the magnetic field. In our model, the magnetic field is merely an external parameter that can be measured point by point via experiment. However, the measurement can be rather difficult. The difficulty is that the magnetic field is a vector field, hence we have to ensure that the surface of the Hall sensor on the Teslameter is precisely parallel to the surface of the magnets, in order to get the correct measurement of the vertical component. Unfortunately, due to the small size of the Hall sensor chip, it is nearly impossible to ensure that it is horizontal within the accuracy we need.

To get around the above problem, we decide to use a theoretical formula of the magnetic field produced by magnet blocks, which can provide us a credible result of the distribution of magnetic field with only a few external parameters that are not hard to obtain. Schlueter and Marks\cite{magnet} have derived the distribution of a magnetic field produced by a rectangular permanent magnet block using the uniform magnetic charge sheet model. For our particular problem, we only consider the case that magnetization is along the y axis. As a result, the charge sheets exist only on the surfaces parallel to the x-z plane, i.e., the normal of the charge sheets are of the same amplitude but opposite signs. For a single charge sheet, whose corners are ${(x_{1},y_{1},z_{1}), (x_{1},y_{1},z_{2}), (x_{2},y_{1},z_{1})}$ and ${(x_{2},y_{1},z_{2})}$, the components of magnetic field are:
\begin{eqnarray}
B_{1x}=\frac{B_{r}}{4\pi}\ln\frac{[z_{2}-z+r(x_{2},y_{1},z_{2})][z_{1}-z+r(x_{1},y_{1},z_{1})]}{[z_{2}-z+r(x_{1},y_{1},z_{2})][z_{1}-z+r(x_{2},y_{1},z_{1})]},\\
\begin{split}
B_{1y}=\frac{B_{r}}{4\pi}\{
\tan^{-1}[\frac{(z_{2}-z)(x_{1}-x)}{(y_{1}-y)r(x_{1},y_{1},z_{2})}]
+\tan^{-1}[\frac{(z_{1}-z)(x_{2}-x)}{(y_{1}-y)r(x_{2},y_{1},z_{1})}]\\
-\tan^{-1}[\frac{(z_{1}-z)(x_{1}-x)}{(y_{1}-y)r(x_{1},y_{1},z_{1})}]
-\tan^{-1}[\frac{(z_{2}-z)(x_{2}-x)}{(y_{1}-y)r(x_{2},y_{1},z_{2})}]\},
\end{split}\\
B_{1z}=\frac{B_{r}}{4\pi}\ln\frac{[x_{2}-x+r(x_{2},y_{1},z_{2})][x_{1}-x+r(x_{1},y_{1},z_{1})]}{[x_{2}-x+r(x_{2},y_{1},z_{1})][x_{1}-x+r(x_{1},y_{1},z_{2})]},
\end{eqnarray}
where,
\begin{equation}
r(x_{i},y_{j},z_{k})=\sqrt{(x_{i}-x)^{2}+(y_{j}-y)^{2}+(z_{k}-z)^2},
\end{equation}
and quantity ${B_{r}}$ is the remanence of the magnet, which can be treated as a fitting parameter in practice. For our magnets, $B_{r}$ is about 600 mT, which leads a field whose vertical component is around 110 mT right on the top surface. (We use the word 'around' here because this parameter is checked every time before an experiment is taken.)

For the other single charge sheet of the block, whose corners are ${(x_{1},y_{2},z_{1}), (x_{1},y_{2},z_{2}), (x_{2},y_{2},z_{1})}$ and ${ (x_{2},y_{2},z_{2})}$, the field components are similar to the previous ones:
\begin{eqnarray}
B_{2x}=-\frac{B_{r}}{4\pi}\ln\frac{[z_{2}-z+r(x_{2},y_{2},z_{2})][z_{1}-z+r(x_{1},y_{2},z_{1})]}{[z_{2}-z+r(x_{1},y_{2},z_{2})][z_{1}-z+r(x_{2},y_{2},z_{1})]},\\
\begin{split}
B_{2y}=-\frac{B_{r}}{4\pi}\{
\tan^{-1}[\frac{(z_{2}-z)(x_{1}-x)}{(y_{2}-y)r(x_{1},y_{2},z_{2})}]
+\tan^{-1}[\frac{(z_{1}-z)(x_{2}-x)}{(y_{2}-y)r(x_{2},y_{2},z_{1})}]\\
-\tan^{-1}[\frac{(z_{1}-z)(x_{1}-x)}{(y_{2}-y)r(x_{1},y_{2},z_{1})}]
-\tan^{-1}[\frac{(z_{2}-z)(x_{2}-x)}{(y_{2}-y)r(x_{2},y_{2},z_{2})}]\},
\end{split}\\
B_{2z}=-\frac{B_{r}}{4\pi}\ln\frac{[x_{2}-x+r(x_{2},y_{2},z_{2})][x_{1}-x+r(x_{1},y_{2},z_{1})]}{[x_{2}-x+r(x_{2},y_{2},z_{1})][x_{1}-x+r(x_{1},y_{2},z_{2})]}.
\end{eqnarray}

The total field is:
\begin{equation}
\vec{B}=\vec{B_{1}}+\vec{B_{2}}.
\end{equation}

The resulting magnetic field, is obtained assuming ${\mu_{r}=1}$, where ${\mu_{r}}$ is the relative permeability. In reality, ${\mu_{r\parallel}=1.04- 1.08}$ and ${\mu_{r\perp}=1.02-1.08}$ \cite{halbach1} \cite{halbach2}. This is the main source of error in the model. Yet our method of calibrating ${B_{r}}$ with measurements removes the error to a large extent, replacing it with measurement error. Of course the anisotropy of the real magnet is still unaccounted for, but it is in general smaller.

Now we have an analytical formula of the magnetic field distribution, but it is still hard to apply to further calculations of eddy currents, since such a complicated formula will appear as a nonhomogeneous term in Equation (14). Therefore, we seek for a simpler formula to replace it. By entering the external parameters of our magnets into the precise formula of the magnetic field, we find that in the region occupied by the gyroscope, the vertical component of the field is solely dependent on coordinate z, as shown in FIG.3.

\begin{figure}[h!]
\centering
\includegraphics[width=5in]{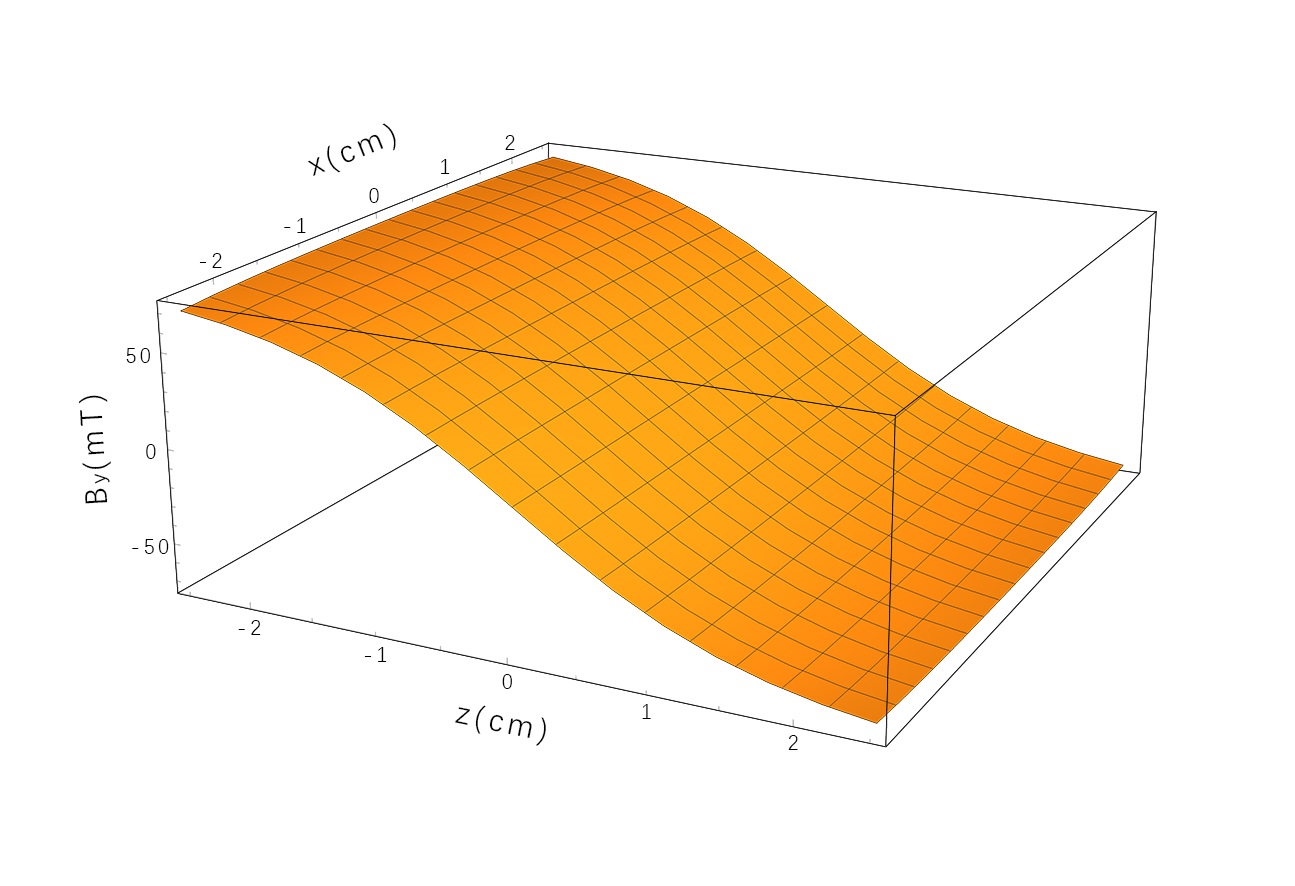}
\caption{The vertical component of the field is almost solely dependent on coordinate z in the region of gyroscope.}
\label{Independence of x}
\end{figure}

Therefore, it is reasonable to describe the magnetic field (vertical component) by a one-parameter function. A conventional way to find a simpler local formula of a function is the method of approximation, which is similar to polynomial curve fitting in experiments. To be specific, we use the following function as an approximation of the field distribution:
\begin{equation}
B_{y}(x,y=h,z) = B_{z} = k_{1}(z-z_{0})+k_{3}(z-z_{0})^{3}.
\end{equation}
where ${k_{1} \&\ k_{3}}$ are undetermined coefficients, and ${z_{0}}$ (which is zero in our model) is the position of the boundary of the two magnets.

Then, with the help of Wolfram Mathematica, the values of ${k_{1}}$ \&\ ${k_{3}}$ are obtained. The result shown in FIG.4 indicates that in the region of the rotator, the analytical and approximate formulas correspond to each other very well.

\begin{figure}[h!]
\centering
\includegraphics[width=5in]{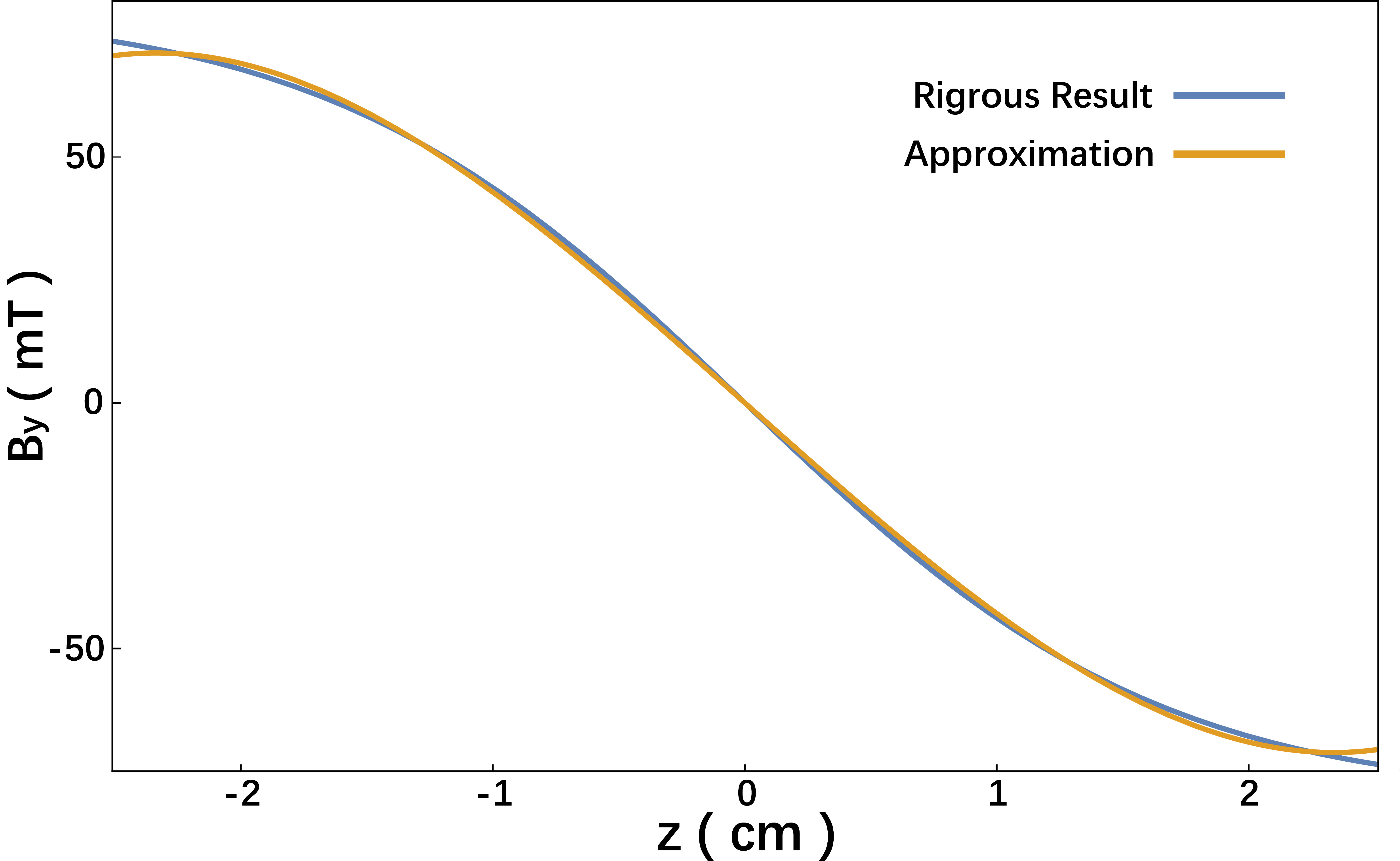}
\caption{The blue line is the result of the original formula, while the orange one is that of Equation(23) with the correct values of ${k_{1} = -45.67 mT\cdot cm^{-1}}$ \&\ ${k_{3} = 2.785 mT\cdot cm^{-3}}$. It can be seen that they fit each other very well.}
\label{Method of approximation}
\end{figure}

\section{Calculation of the current}
Based on this preliminary work, we can transform our 'gyroscope teslameter' problem into a boundary value problem of Poisson's equation:
\begin{eqnarray}
\begin{cases}
\nabla^{2}\varphi = \vec\nabla\cdot[(\vec{\omega}\times\vec{r})\times\vec{B}],\\
j_{n}\mid_{boundary} = \sigma(\vec E + \vec f)\cdot\hat{n}\mid_{boundary} = 0.
\end{cases}
\end{eqnarray}

Since the rotator of our gyroscope is not simply a cylinder, but a combination of two rings and a circular plate (see FIG.5), we decide to solve the Poisson's equation in the rings and the plate separately, and then combine them together. As a result, the problem in each region reduces to 2-dimensions (since they are translational invariant along the y axis). Therefore, we solve the following equations:
\begin{eqnarray}
\begin{cases}
\nabla^{2}\varphi = \vec\nabla\cdot[\omega r (k_{1}r \sin\theta + k_{3}r^{3}\sin^{3}\theta)\hat{r}] \\
\quad\quad\,= \omega [3 k_{1} r \sin\theta + \frac{5}{4} k_{3} r^{3}(3\sin\theta-\sin 3 \theta)],\\
\frac{\partial\varphi}{\partial r}\mid_{r=a} - \omega a (k_{1}a \sin\theta + k_{3}a^{3}\sin^{3}\theta)= 0,\\
\frac{\partial\varphi}{\partial r}\mid_{r=b} - \omega b (k_{1}b \sin\theta + k_{3}b^{3}\sin^{3}\theta)= 0,
\end{cases}
\end{eqnarray}
where ${a}$ and ${b}$ are inside and outside diameter of the ring. For the plane region, we just take ${a=0}$. Here we make use of cylindrical coordinates (or polar coordinates since the problem is 2D) them. The relations between them and the original cartesian coordinates are:
\begin{eqnarray}
\begin{cases}
x = r\cos\theta,\\
y = y,\\
z = r\sin\theta.
\end{cases}
\end{eqnarray}

\begin{figure}[h!]
\centering
\includegraphics[width=2in]{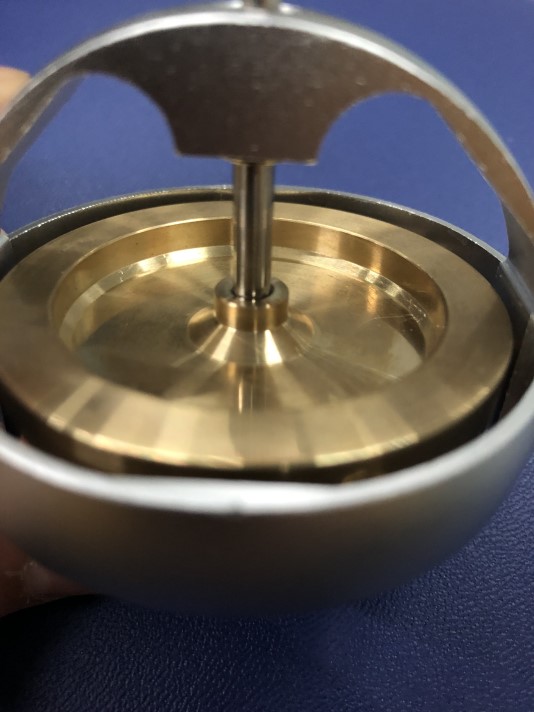}
\caption{This is a close-up view of the gyroscope. It can be seen that the rotator is a combination of two rings and a plate (there is only one ring can be seen on the picture, and the other one is on the bottom side).}
\label{Shape of the rotator}
\end{figure}

To solve the Equation (25), we use the superposition principle to split this Poisson's equation with the complicated boundary conditions into two equations. One of these equations is the same Poisson's equation with all boundary conditions equal to zero, i.e. a inhomogeneous partial differential equation with homogeneous boundary conditions:
\begin{eqnarray}
\begin{cases}
\nabla^{2}\varphi_{P} = \omega [3 k_{1} r \sin\theta + \frac{5}{4} k_{3} r^{3}(3\sin\theta-\sin 3 \theta)],\\
\frac{\partial\varphi_{P}}{\partial r}\mid_{r=a}= 0,\\
\frac{\partial\varphi_{P}}{\partial r}\mid_{r=b}= 0.
\end{cases}
\end{eqnarray}
This equation can be solved by assuming that the solution is a summation of infinite terms: ${\varphi_{P} = \sum_{n=0}^{\infty}\varphi_{n}}$, where ${\varphi_{n} = A_{n}(r)\cos(n\theta)+B_{n}(r)\sin(n\theta)}$. By plugging it into the Equation (27), the partial differential equation is reduced to second order ordinary differential equations of variable ${r}$, which are named 'Euler's equation' in mathematical physics, each associated with one function: ${A_{n}(r)}$ or ${B_{n}(r)}$ with a particular ${n}$. There is a systematic way to solve Euler's equation, and Equation (27) is then solved. It is worth noticing that the form of solution we assumed above is not arbitrary, but is based on the fact that ${\cos(n\theta)}$ and ${\sin(n\theta)}$ with integer ${n}$ are the 'angular eigenfunctions' of Poisson's equation (and Laplace's equation) in such a region.

The other equation (Equation 27), then becomes a Laplace's equation with the same boundary conditions of the initial equation (25), i.e. a homogeneous equation with inhomogeneous boundary condition:
\begin{eqnarray}
\begin{cases}
\nabla^{2}\varphi_{L} = 0,\\
\frac{\partial\varphi_{L}}{\partial r}\mid_{r=a} = \omega a (k_{1}a \sin\theta + k_{3}a^{3}\sin^{3}\theta),\\
\frac{\partial\varphi_{L}}{\partial r}\mid_{r=b} = \omega b (k_{1}b \sin\theta + k_{3}b^{3}\sin^{3}\theta),
\end{cases}
\end{eqnarray}
which can be solved the same way as above.

The solutions ${\varphi_{P}}$ and ${\varphi_{L}}$ to these two equations can be added to get the solution ${\varphi}$ to the initial equation:
\begin{equation}
\varphi = \varphi_{P} + \varphi_{L}.
\end{equation}

Finally, we obtain an analytical result of the scalar potential distribution. Subsequently, we can get the electric field by evaluating ${\vec E = -\vec\nabla \varphi}$, and then plug ${\vec E}$ back into Equation (5) to get the current distribution. Since the final formula is rather long, we list the current induced by the first and second term of the magnetic field (see Equation 23) separately, naming them ${\vec j_{1}}$ and ${\vec j_{3}}$ respectively:
\begin{equation}
\begin{aligned}
\begin{split}
\vec j_{1}(r,\theta)= &\sigma\omega k_{1} [r^{2}\sin\theta+\frac{1}{8}(a^{2}+b^{2}-\frac{a^{2}b^{2}}{r^{2}}-9r^{2})\sin\theta] \hat{r}\\
&+\sigma\omega k_{1}[\frac{1}{8}(a^{2}+b^{2}-\frac{a^{2}b^{2}}{r^{2}}-3r^{2})\cos\theta] \hat{\theta},
\end{split}\\
\end{aligned}
\end{equation}
\begin{equation}
\begin{aligned}
\begin{split}
\vec j_{3}(r,\theta)= &\sigma\omega k_{3}\frac{(a^{2}-r^{2})(r^{2}-b^{2})}{64(a^{4}+a^{2}b^{2}+b^{4})r^{4}}[2(a^{4}+a^{2}b^{2}+b^{4})(a^{2}+b^{2}+r^{2})r^{2}\sin\theta\\
&-9(r^{4}(a^{4}+a^{2}b^{2}+b^{4})+r^{2}a^{2}b^{2}(a^{2}+b^{2})+a^{4}b^{4})\sin3\theta] \hat{r}\\
&+ \frac{\sigma\omega k_{3}}{64r^{4}}[2r^{2}(a^{2}b^{2}(a^{2}+b^{2})+r^{2}(a^{4}+a^{2}b^{2}+b^{4})-5r^{6})\cos\theta\\
&-\frac{9r^{6}(a^{6}+a^{4}b^{2}+a^{2}b^{4}+b^{6})-15r^{8}(a^{4}+a^{2}b^{2}+b^{4})+9a^{6}b^{6}}{a^{4}+a^{2}b^{2}+b^{4}}\cos3\theta] \hat{\theta}.
\end{split}
\end{aligned}
\end{equation}

The above two formulas represent the current distributions in the rotator analytically. To give a more intuitive picture, we use a built-in function of Wolfram Mathematica to generate the following stream plots of the currents (see FIG.6, the formulas of boundary conditions are also listed there). These stream plots are slightly different from the conventional ones that illustrate electric and magnetic fields, since the density of curves here does not represent the intensity of current. Only the directions of arrows have physical significance as the direction of current at each point.

\begin{figure}[h!]
\centering
\subfigure[ ]{
\includegraphics[width=5.5cm]{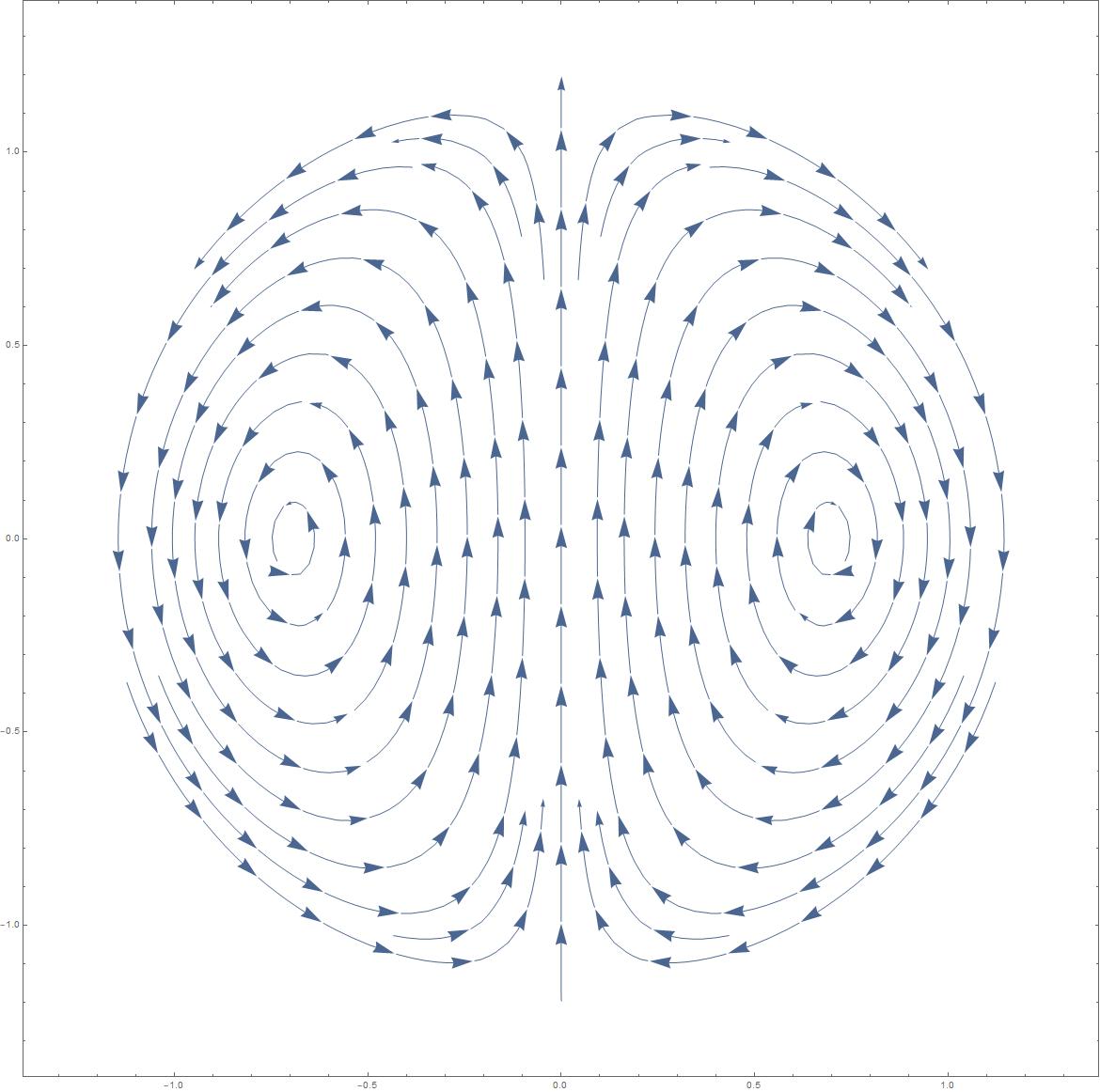}
}
\quad
\subfigure[ ]{
\includegraphics[width=5.5cm]{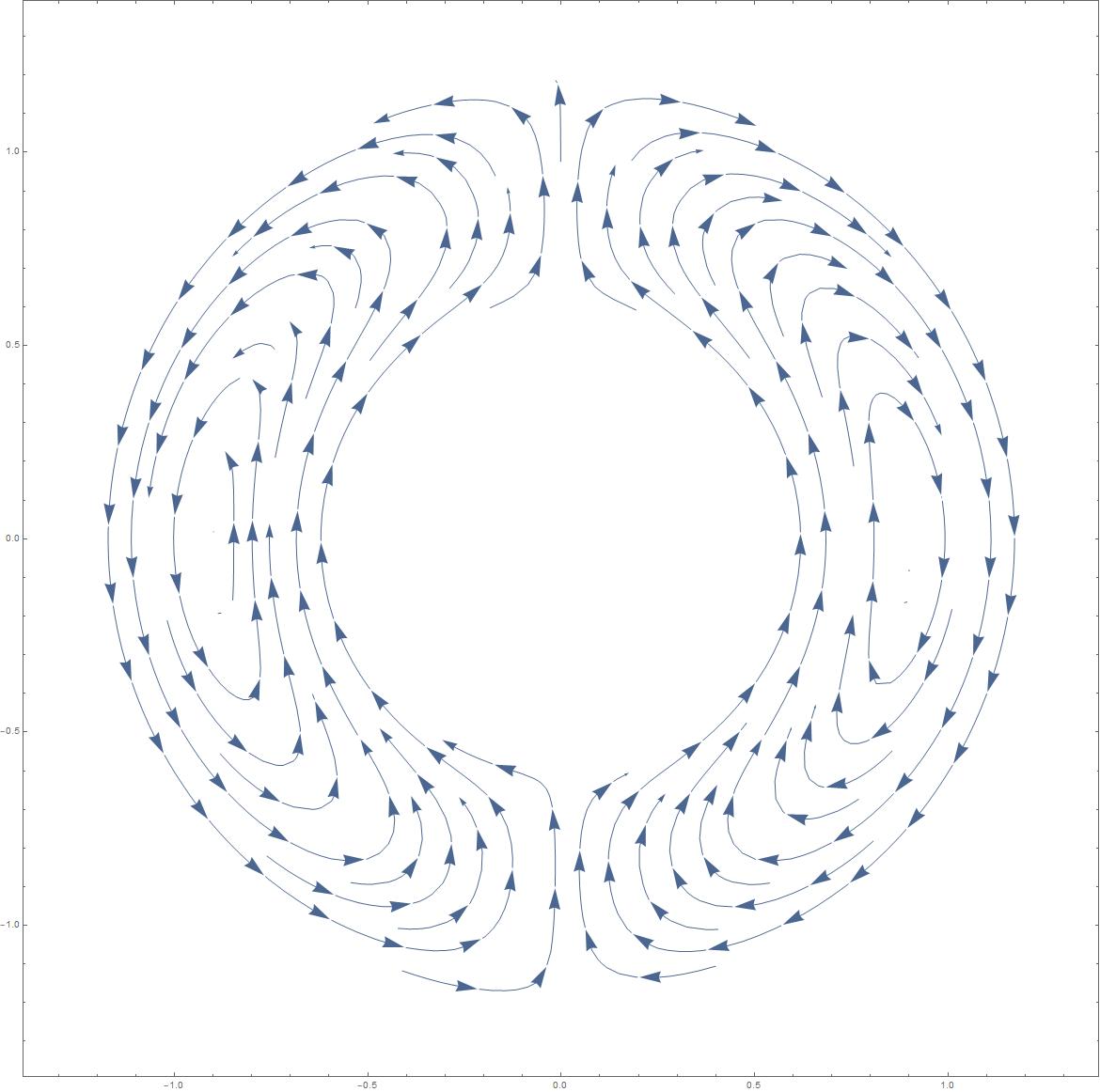}
}
\quad
\subfigure[ ]{
\includegraphics[width=5.5cm]{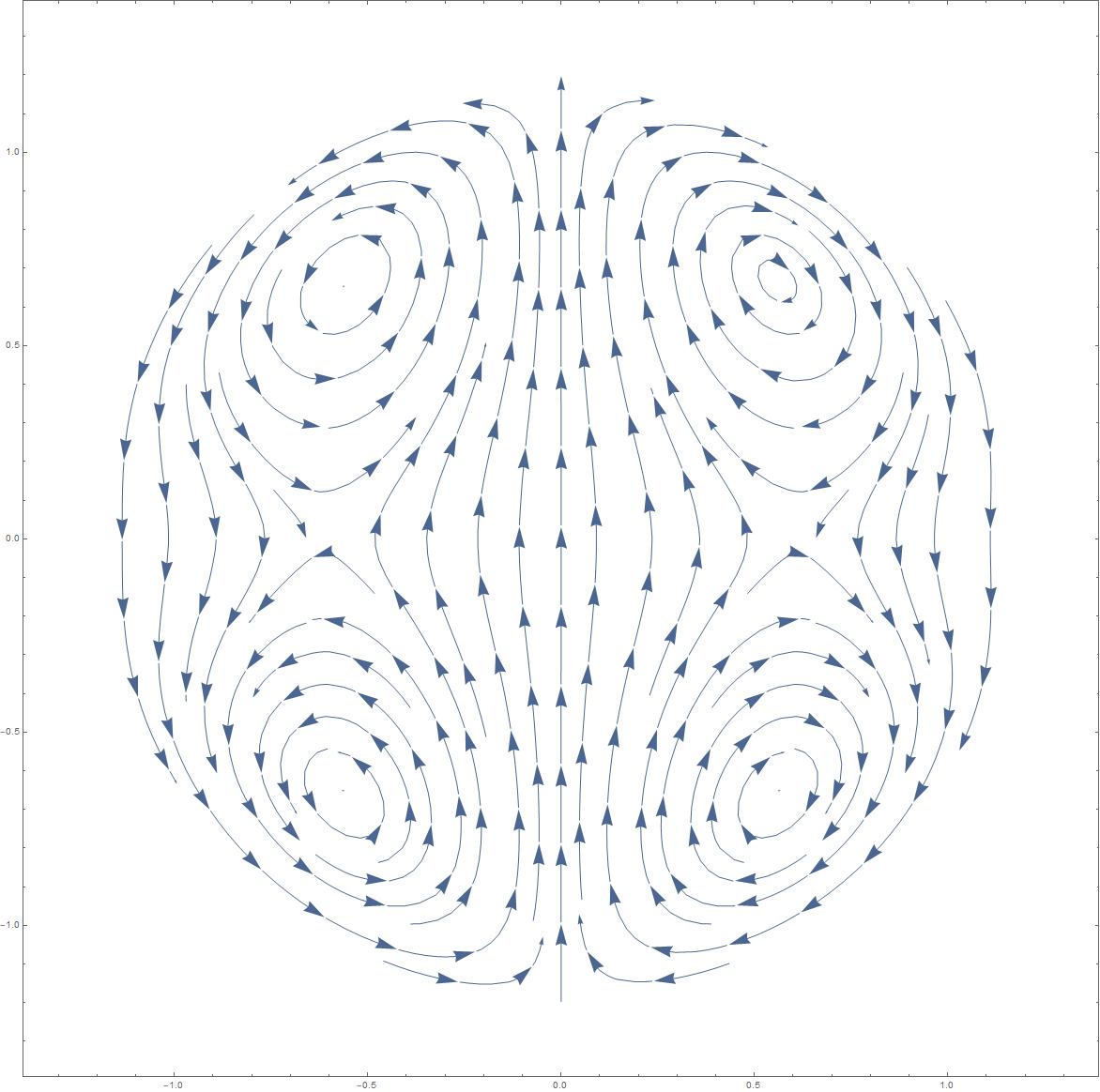}
}
\quad
\subfigure[ ]{
\includegraphics[width=5.5cm]{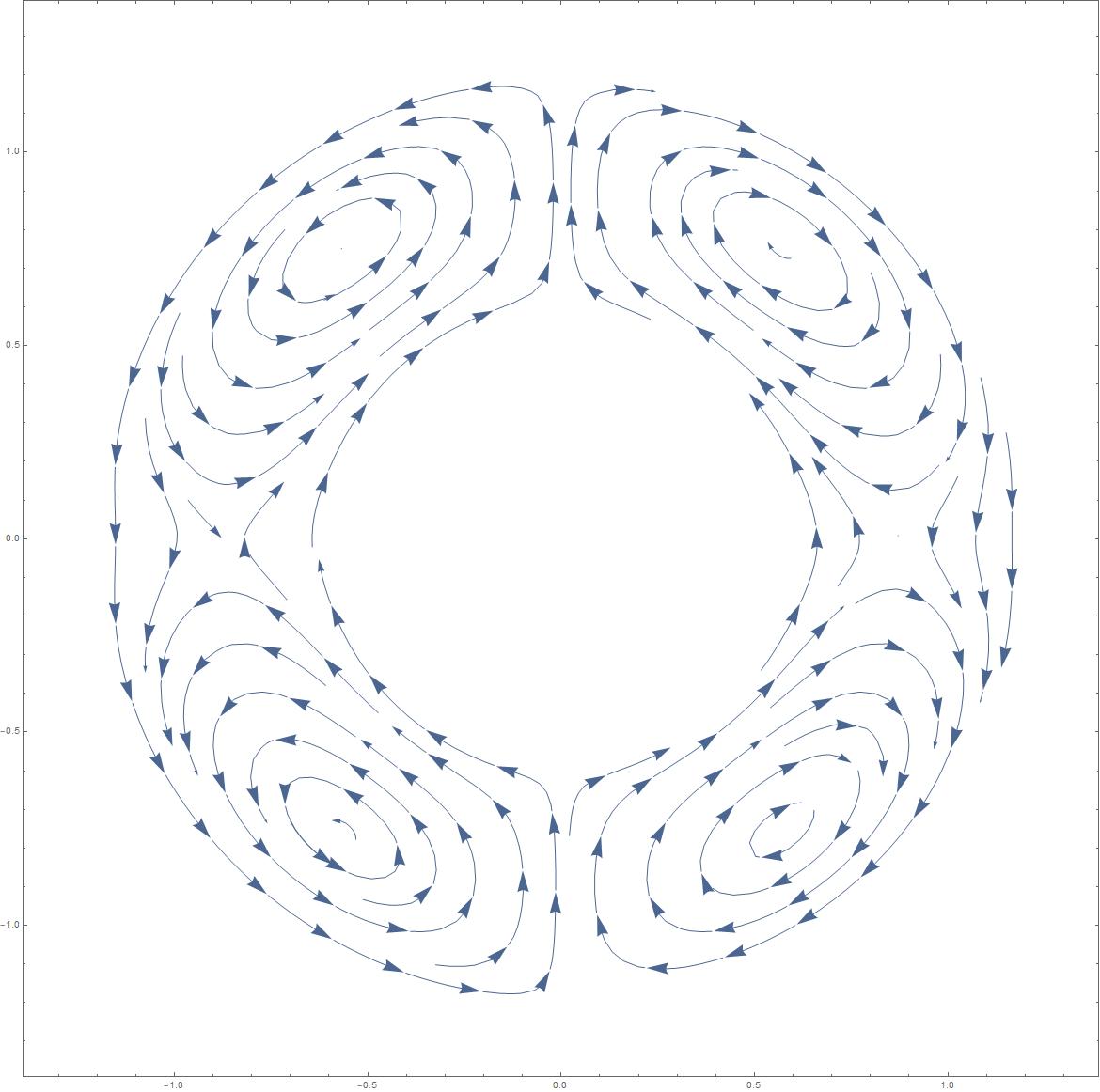}
}
\caption{These are stream plots of the currents generated by Wolfram Mathematica. Figure (a) represents eddy current in a circle plate (or a cylinder) induced by a 'linear magnet field', that is, the first term of Equation (23) (Equation(30)). Figure (b) represents eddy current in a ring induced by the same field. And figures (c) and (d) show the currents induced by a 'cubic magnetic field', that is, the second term of Equation (23) (Equation(31)). One interesting feature of these plots is the degree of symmetry. There are left-right symmetry and top-down anti-symmetry, which reflects the symmetry of the magnetic field.}
\label{Eddy current}
\end{figure}

\section{Comparison with experiments}
With the formulas of current distribution, Equation (30) \&\ (31), we can easily calculate the magnetic torque ${\vec M}$ on the rotator of gyroscope via the following integral:
\begin{equation}
\vec M = \int_{V}\vec{j}\times\vec{B} dv = -\lambda\vec\omega,
\end{equation}
where coefficient ${\lambda}$ is independent of angular velocity. That the torque is proportional to angular velocity is obvious since the current in the integral is proportional to it as well. The following law of mechanics is taken here:
\begin{equation}
I\frac{d\omega}{dt} = M = -\lambda\omega.
\end{equation}
In the solution
\begin{equation}
\omega (t) = \omega_{0} e^{-t / \tau},
\end{equation}
${\tau = I/\lambda}$ is the 'characteristic decaying time'.

Entering the dimensions and the conductivity of our gyroscope, we get the result that ${\tau = 7.839 s}$.

On the other hand, the measurement of deceleration of the rotator above magnets is done with the favor of a strobe light (see FIG.2 for details). However, we realize that the experimental result is not an exponential decay, but with a minus constant added, as shown below:
\begin{equation}
\omega (t) = \omega_{0} e^{-t / \tau_{exp}} - \omega_{rest},
\end{equation}
where the subscript 'exp' (refers to 'experiment') is to distinguish from the theoretical one.

\begin{figure}
  \centering
  \includegraphics[width=5in]{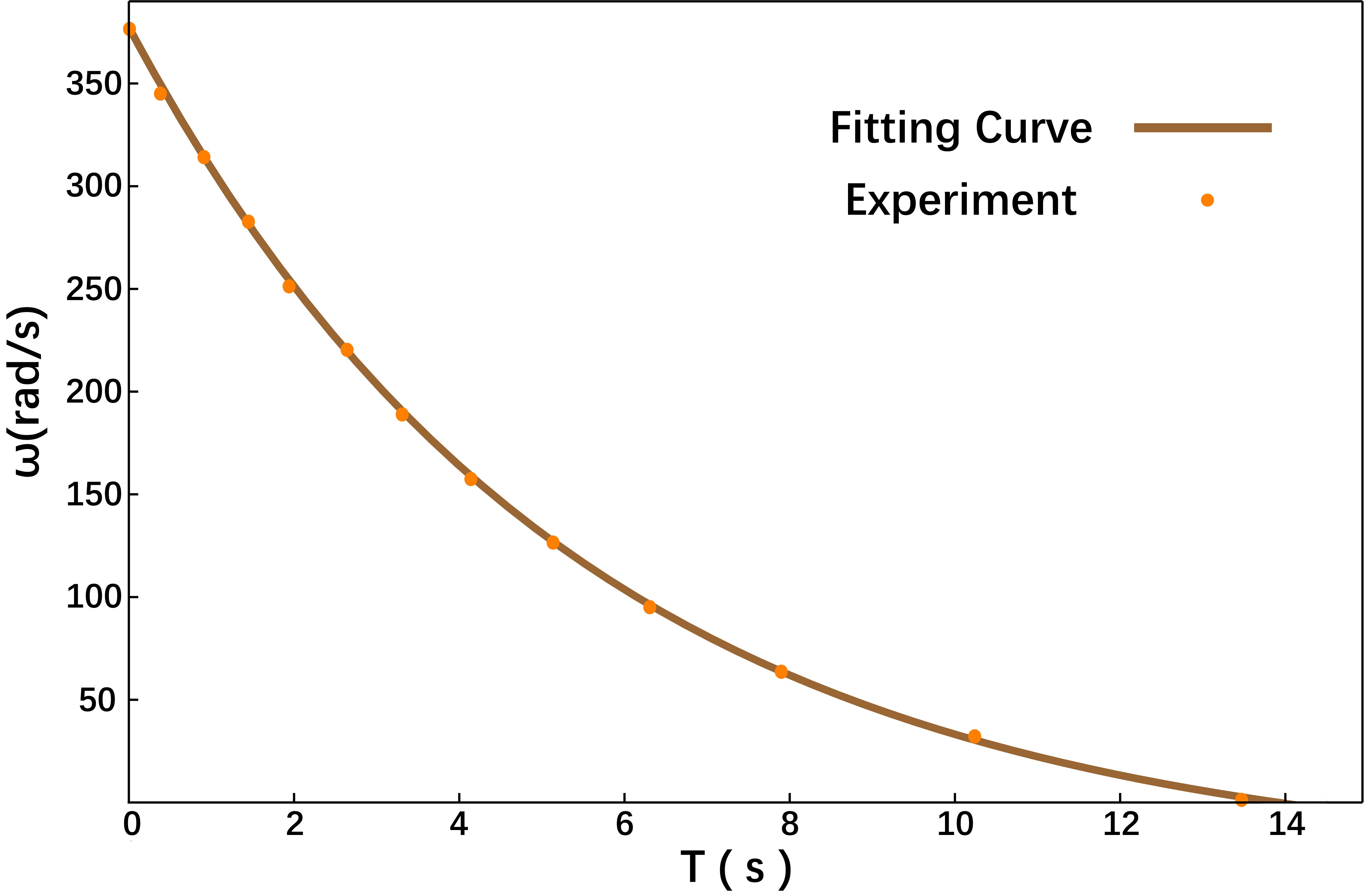}
  \caption{Measurement of the deceleration process, which is displayed as orange points, is fit by a curve (brown) with the form of Equation (35). In our experiments, the initial angular velocity is ${\omega_{0} - \omega_{rest}= 377.0 rad/s}$, which is 60.00 Hz, and the constant term is ${\omega_{rest} = 32.57 rad/s}$. The result of decaying time (given by the fitting function of OriginLab) is ${\tau_{exp} = 5.420s}$. Moreover, the correlation coefficient is ${r = 0.99993}$, which indicates that Equation (35) is precise enough to represent the deceleration process. \\
  Notice: we do not draw error bars on the points since the error caused by measurement is too small to be shown on the plot.}\label{Fitting curve}
\end{figure}

Such a result is reasonable since an ideal exponential decay indicates that the gyroscope will not stop spinning forever, which is unrealistic. After examining Equation (33) carefully, we found that if a constant frictional force is added to the right hand side, a decelerating curve of the same form with the experimental one is obtained.

Consequently, we turn to investigate the frictional force here. Through experiments, we measure the deceleration curve of the gyroscope without the magnets, which means the only torque on the rotator is caused by frictional force. Such a deceleration curve gives directly the form of the torque generated by the frictional force, which is linear to the angular velocity:
\begin{equation}
M_{f} = -M_{0}-\alpha \omega.
\end{equation}

However, this is not the end of the story. An intrinsic shortcoming of our gyroscope is that its shaft is made of steel, which is magnetic. The attraction between the shaft and the magnets exerts additional pressure between the rotator and the frame which enlarges frictional force. Therefore, the frictional torque we measured away from the magnets is not identical to that in the presence of the magnets.
Fortunately, the empirical formula of frictional force: ${F_{f} = \mu F_{N}}$ indicates that it is proportional to pressure, hence the torque we are looking for is likely to be a multiple of the one measured in absence of the magnets. In our model, we treat this multiplier as a fitting parameter. By choosing it properly, we are able to make the constant term in the theoretical deceleration curve the same as that of the experimental curve. In the end, we achieve good agreement between the theory and experiments (see FIG.8), the error in the characteristic decaying time is about 8.61\%\ .

\begin{figure}[h!]
\centering
\includegraphics[width=5in]{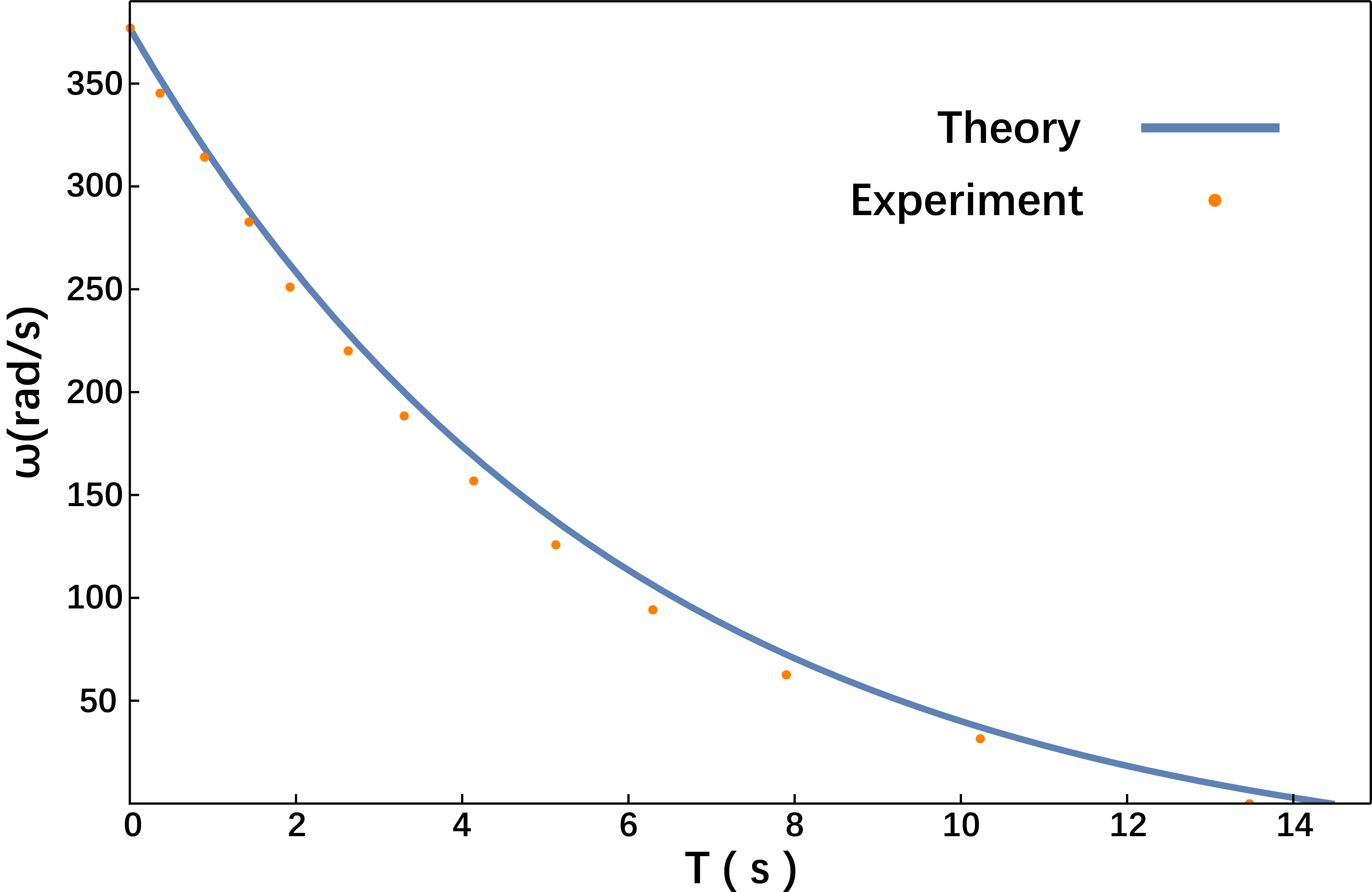}
\caption{A comparison between the theory and the experiment. As stated below FIG.7, the value of decaying time of our experiment is ${\tau_{exp} = 5.420 s}$. Our calculation of the interaction between magnetic field and eddy current, together with the calibration of the frictional torque, gives that with the same initial conditions, the decaying time is ${\tau = 5.887 s}$ . The error in the characteristic decaying time is about 8.61 \%\ .}
\label{Final result}
\end{figure}

\section{Reasonability of the assumptions}
In this section, we give a brief review of the assumptions we introduced in Section II and discuss possible sources of error.

The first assumption we introduced is that we treat the current (as well as electric field) as nearly time-independent. As all the fields, charge and current distribution varying 'slowly�as time goes on, quantity ${\partial \rho / \partial t}$ will be small enough to be omitted, hence the divergence of the current is zero. To be more specific, the condition of nearly time-independent that can be expressed as the partial derivative of electric field (or current) is much less than that of electric field itself (of course with some constants to scale them to the same dimension):\cite{cai}
\begin{equation}
{\varepsilon\frac{\partial\vec{E}}{\partial t}} \ll \sigma\vec{E}.
\end{equation}
As for our brass rotator, ${\frac{\sigma}{\epsilon} = 1.613\times 10^{18} s^{-1}}$, which has a huge dimension so that the electric field and the current can be treated as time-independent.

Our second assumption is straight forward, and is what people usually take when calculating dynamic electromotive force.\cite{Nurge2018}

Regarding the third assumption, which is that the magnetic field gives a force to the current, and this force acts on the gyroscope directly to make it slow down without affect current distribution, is justified by the fact that under normal circumstance, where electrons move at non-relativistic velocity electric force on a moving charge will be much more significant than magnetic force, hence the distribution of the current will be determined primarily by the electric field. Being a higher order perturbation, magnetic force will drag the moving electrons aside slightly, but such a transversal movement will be prevented by rapid collision between electrons and lattice. As a result, the magnetic force is transfer to lattice of the conductor.

Finally, we would like to discuss a possible source of the 8.61 \%\ error in ${\tau}$. Besides all the approximations we made in process of calculation, the most significant origin of the error is that the magnetic field is not actually vertical. In fact, there is a strong horizontal component of the same dimension with the vertical one, which may also induce a vertical component of current and then another torque. Due to the fact that the thickness of the gyroscope is small relative to its diameter, the energy loss is small compared to the main effect. However, such a current is not easy to calculate, so only a limit of error is here given by coarse estimation, which is about 10 \%\ in torque and also in characteristic decaying time.

\section{Conclusion}
The piece of work introduced here provides an analytical method to obtain the distribution of the eddy current in a thin metal disk or ring and the
damping time of the rotation, whose validity has been verified by experiments. Compared to numerical simulation, this method has the advantage of
much easier access and shorter computation time. Neither specialized professional software package nor high performance computers are necessary.
Despite the fact that the configuration of the magnetic field used in the paper is non-generic, the result presented in the paper can be readily applied
to arbitrary distribution of magnetic field. One only has to decompose the magnetic field component perpendicular to the disk in its vicinity into
symmetric and anti-symmetric parts and then follow steps in the paper to obtain the damping time. In case that analytical expression of the magnetic
field distribution is not available, measurement or numerical data can be used. After the decomposition is done, a third-order polynomial fit can be
done. Again, the damping time can be calculated based on the method of this paper.

We hope the work can assist students around the world to learn the eddy currents and give them a deeper understanding of Maxwell equations and how to apply them in real cases. We intentionally omit many details of solving the Poisson's equation since we do not want the main logic of physics to be obstructed by mathematical details.

\begin{acknowledgments}
Our program is strongly supported by School of Physical Science and Technology at ShanghaiTech University. The authors would like to thank Dr. Yintao You for giving a feasible plan to measure the angular velocity. Especially, the authors should appreciate Prof. Ramamurti Shankar from Yale University who helped to revise the whole paper carefully, as well as Ms. Bingrui Wu from College of Physics, Jilin Uiversity, who helped us a great deal in making use of computer programs to do calculation, designing the figures to better illustrate the results, and editing this article with LATEX. In addition, we wish to thank Dr. Zhongkai Liu, Dr. Jiamin Xue, Dr. Fuxiang Han, and Dr. Xiaoping Liu for many useful discussions and advice, and Mrs. Kangning Wang and Dean Xuguang Xu for helping us contact with different departments in the university for support.
\end{acknowledgments}

\end{document}